# Spatial Queries with Two kNN Predicates[*]

Ahmed M. Aly
Purdue University
West Lafayette, IN
aaly@cs.purdue.edu

Walid G. Aref
Purdue University
West Lafayette, IN
aref@cs.purdue.edu

Mourad Ouzzani
Qatar Computing Research Institute
Qatar
mouzzani@qf.org.qa

## ABSTRACT

The widespread use of location-aware devices has led to countless location-based services in which a user query can be arbitrarily complex, i.e., one that embeds multiple spatial selection and join predicates. Amongst these predicates, the $k$-Nearest-Neighbor ($k$NN) predicate stands as one of the most important and widely used predicates. Unlike related research, this paper goes beyond the optimization of queries with single $k$NN predicates, and shows how queries with *two* $k$NN predicates can be optimized. In particular, the paper addresses the optimization of queries with: (i) two $k$NN-select predicates, (ii) two $k$NN-join predicates, and (iii) one $k$NN-join predicate and one $k$NN-select predicate. For each type of queries, conceptually correct query evaluation plans (QEPs) and new algorithms that optimize the query execution time are presented. Experimental results demonstrate that the proposed algorithms outperform the conceptually correct QEPs by orders of magnitude.

## 1. INTRODUCTION

Many emerging applications of location-based services demand complex location-based queries. These queries can contain multiple predicates that involve a combination of spatial (e.g., $k$NN and range) predicates along with the traditional selects, joins, and group-by's of relational databases.

Although a large spectrum of research has been devoted to query processing of location-based queries (e.g., [12, 5, 10, 11, 21, 20, 9, 8]), *none* addresses the processing and optimization of location-based queries that contain multiple location-based predicates.

The key issue in queries with multiple location-based predicates is that they can produce different results based on the order in which the predicates are evaluated. This results in an ambiguity on the intended semantics of these queries. In [19], we study the conceptual evaluation of queries that include multiple similarity predicates [16]: similarity group-by (e.g., group-around) [17], similarity join (e.g., $\epsilon$-join, $k$NN-join, and join-around) [18], and similarity selection (e.g., $\epsilon$-selection and $k$NN-selection). In [19], we provide equivalence rules for similarity queries in the form of algebraic transformations that focus on the *correctness* of these transformations, but do not introduce any algorithms for the efficient evaluation of similarity queries. In contrast, this paper introduces efficient algorithms for processing queries with two $k$NN predicates while retaining the correctness of their evaluation.

In this paper, we focus on the operations: $k$NN-select and $k$NN-join. While these operations have a variety of flavors, the ones we adopt in this paper are explained as follows. Assume that we have two sets, say $E_1$ and $E_2$, of points in the two-dimensional space. For simplicity, we use the Euclidean distance.

- **$k$NN-select**: For a focal point $f$, $\sigma_{k,f}(E_1)$ returns from the set of points in $E_1$ the $k$-closest to $f$.

- **$k$NN-join**: $E_1 \bowtie_{kNN} E_2$ returns all the pairs of the form $(e_1, e_2)$, where $e_1 \in E_1$ and $e_2 \in E_2$, and $e_2$ is among the $k$-closest points to $e_1$.

Queries containing two of these operations embed significant query processing and optimization challenges. For example, the well-known heuristic of pushing selections below joins [4] to reduce the execution time of a query, can produce wrong results in the case of a $k$NN-join. This is demonstrated through the following example.

Assume that a car breaks while in travel. The driver needs to find an hotel and a mechanic shop that are close to each other. At the same time, the driver wants the hotel to be close to a specific shopping center, so that he can do shopping while the car is being repaired. The driver issues the following query: From the list of mechanic shops and the two closest hotels to each mechanic shop, report the (mechanic shop, hotel) pairs, where the hotel is amongst the two closest neighbors of the shopping center.

Notice that this query involves a $k$NN-select on the inner (right) relation of a $k$NN-join. Figures 1 and 2 give two possible QEPs for the query. In both figures, black dots represent mechanic shops, white dots represent hotels, and the red triangle represents the shopping center. In Figure 1, the $k$NN-select is performed after the $k$NN-join, while in Figure 2, the $k$NN-select is pushed below the $k$NN-join. As the figures demonstrate, the two QEPs produce different results.

According to [19], the correct QEP for such query is the one in Figure 1. Pushing a $k$NN-select under the inner relation of a $k$NN-join; as a standard relational query optimizer would typically do; reduces the scope of the points being considered in the inner relation. When the $k$NN-join is performed, the outer relation will not have the entire set of points of the inner relation to join with, and hence, the $k$NN-join will not be performed correctly. For example, in Figure 2, the Mechanic relation will have nothing to join with except Hotels $h_1$ and $h_2$. Thus, the resulting pairs will be *all* the mechanic shops paired with either $h_1$ or $h_2$, which is wrong. In other words,

$$(E_1 \bowtie_{kNN} E_2) \cap (E_1 \times \sigma_{k_\sigma, f}(E_2)) \not\equiv E_1 \bowtie_{kNN} (\sigma_{k_\sigma, f}(E_2)).$$

[*]This work was partially supported by the National Science Foundation under Grants III-1117766, IIS-0964639, and IIS-0811954.

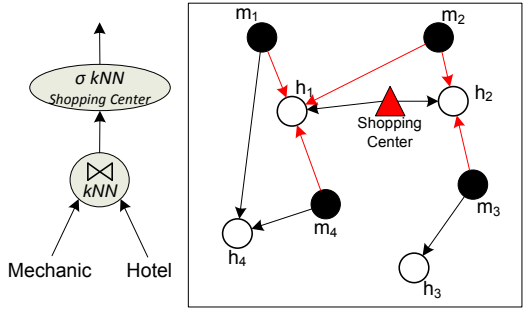

**Figure 1: A QEP with the $k$NN-select performed after the $k$NN-join. $k = 2$ in both predicates. The resulting pairs are: $(m_1, h_1)$, $(m_2, h_1)$, $(m_2, h_2)$, $(m_3, h_2)$, and $(m_4, h_1)$.**

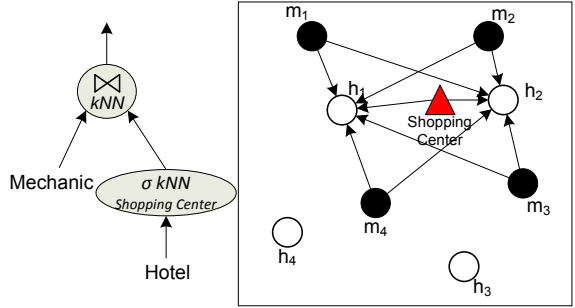

**Figure 2: A QEP with the $k$NN-select pushed below the inner relation of the $k$NN-join. $k = 2$ in both predicates. The resulting pairs are: $(m_1, h_1)$, $(m_1, h_2)$, $(m_2, h_1)$, $(m_2, h_2)$, $(m_3, h_1)$, $(m_3, h_2)$, $(m_4, h_1)$, and $(m_4, h_2)$.**

The above example demonstrates that pushing a $k$NN-select on the inner relation of a $k$NN-join is invalid. The lack of such optimization calls for new optimization techniques that can still leverage the pruning effect of selection without compromising the correctness of evaluation.

In addition to the above form of interaction between $k$NN predicates, we study the following cases:

- The case of a $k$NN-select on the outer relation of a $k$NN-join. This case has been added for completeness. Actually, pushing a selection below the outer relation of a $k$NN-join produces correct query results.

- The cases of two *chained* and *unchained* $k$NN-joins. Since the $k$NN-join is not a symmetric operation, the two expressions $(E_1 \bowtie_{kNN} E_2) \cap (E_2 \bowtie_{kNN} E_3)$ and $(E_1 \bowtie_{kNN} E_2) \cap (E_3 \bowtie_{kNN} E_2)$ are not equivalent. We call the joins in the former expression chained ($E_1 \rightarrow E_2 \rightarrow E_3$), and those in the latter expression unchained.

- The case of two $k$NN-selects.

For each of these cases, we introduce efficient algorithms that not only guarantee the correctness of evaluation, but also outperform the corresponding conceptually correct QEPs by orders of magnitude.

More specifically, the contributions of this paper can be summarized as follows.

1. We introduce two algorithms for evaluating a query with a $k$NN-select on the inner relation of a $k$NN-join (Section 3).

2. We study the cases of two chained and unchained $k$NN-joins, and introduce efficient algorithms for their evaluation (Section 4).

3. We study the case of two $k$NN-selects, and present an efficient algorithm for its evaluation (Section 5).

4. We conduct extensive experiments that show how our proposed techniques outperform the conceptually correct QEPs by orders of magnitude (Section 6).

## 2. PRELIMINARIES

We assume that the data consists of points in the two-dimensional space. The algorithms we present do not assume a specific indexing structure. The algorithms can be applied to a quadtree, an R-tree, or any of their variants (e.g., [14, 6, 2, 7]). The quadtree and its variants are hierarchical spatial data structures that recursively partition the underlying space into blocks until the number of points inside a block satisfies some criterion (being less/greater than some threshold). We assume that the index maintains the *count* of points in each block. We use a simple grid in the figures for illustration purposes.

In this paper, we make extensive use of the two metrics: MINDIST and MAXDIST [13]. The MINDIST (or MAXDIST) between a point, say $p$, and a block, say $b$, refers to the minimum (or maximum) possible distance between $p$ and any point in $b$. In the algorithms we present, we process the blocks in a certain order according to their MINDIST (or MAXDIST) from a certain point. An ordering of the blocks based on the MINDIST or MAXDIST from a certain point is termed a MINDIST or MAXDIST ordering, respectively. We use the terms: *neighborhood* and *locality* of a point [15] that are defined as follows:

DEFINITION 1. The **neighborhood** of a point, say $p$, is the set of the $k$ nearest neighboring points to $p$.

DEFINITION 2. The **locality** of a point, say $p$, is a set of blocks inside which the neighborhood of $p$ exists.

One can use any algorithm to compute the neighborhood of a point. In this paper, we employ the *locality* algorithm of [15]. Given a point, say $p$, the main idea of the algorithm is to build the minimum locality of $p$, and then compute the neighborhood of $p$ only from its locality. For more detail on the algorithm, the reader is referred to [15].

## 3. KNN-SELECT WITH KNN-JOIN

As discussed in Section 1, pushing a $k$NN-select on the inner relation of a $k$NN-join is invalid. However, pushing a $k$NN-select on the outer relation of a $k$NN-join is valid [19], i.e.,

$$(E_1 \bowtie_{kNN} E_2) \cap ((\sigma_{k_\sigma, f}(E_1)) \times E_2) \equiv (\sigma_{k_\sigma, f}(E_1)) \bowtie_{kNN} E_2.$$

To illustrate the above situation, consider the scenario in Section 1. Assume that the driver issues the following query: From the list of mechanic shops and the two closest hotels to each mechanic shop, report the (mechanic shop, hotel) pairs where the mechanic shop is amongst the two closest neighbors of the shopping center.

Notice that in this case, the selection is on the outer (left) relation of the join.

Figure 3 gives two different QEPs; $QEP_1$ and $QEP_2$; for the query. In $QEP_1$, the selection is pushed below the join while in $QEP_2$, the selection is performed after the join. Clearly, both QEPs produce the same results. This is because as a consequence of the pushed selection in $QEP_1$, some points of the outer relation will be excluded from the join. However, performing the join for these excluded points is useless as the results of the join that have any of these points will have to be excluded anyway if the selection is to be applied at the end, as in $QEP_2$.

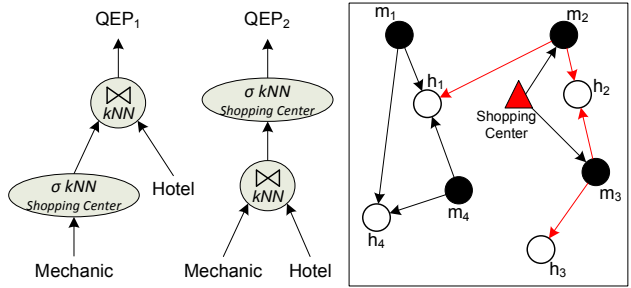

Figure 3: Two QEPs for a query with a $k$NN-select on the outer relation of a $k$NN-join. $k = 2$ in both predicates. Both QEPs result in the same pairs: $(m_2, h_1)$, $(m_2, m_2)$, $(h_3, h_2)$, and $(m_2, h_3)$.

The challenge in pushing a $k$NN-select[1] on the inner relation of a $k$NN-join calls for new optimization techniques that can still leverage the pruning effect of selection without compromising the correctness of evaluation.

In the rest of this section, we present two algorithms; **Counting** and **Block-Marking**; for evaluating a query with a $k$NN-select on the *inner* relation of a $k$NN-join. Formally, the two algorithms evaluate a query of the form $(E_1 \bowtie_{kNN} E_2) \cap (E_1 \times \sigma_{k_\sigma, f}(E_2))$, that retrieves the pairs $(e_1, e_2)$, such that $e_2$ is $k_\bowtie$-closest to $e_1$ and $k_\sigma$-closest to $f$, where $k_\bowtie$ is the $k$ value of the join, and $k_\sigma$ is the $k$ value of the selection.

The two algorithms are based on the following insight. First, we compute the neighborhood of $f$ (i.e., perform the selection). Then, for each point $e_1 \in E_1$, if we can make sure that the neighborhood of $e_1$ *cannot* intersect the neighborhood of $f$ *without* computing the neighborhood of $e_1$, then we ignore $e_1$ as it will not contribute to the results of the query. Otherwise, we compute the neighborhood of $e_1$, and intersect it with the neighborhood of $f$. The difference between the two algorithms is in the way they check if the neighborhood of $e_1$ cannot intersect the neighborhood of $f$.

### 3.1 Counting Algorithm

The Counting algorithm proceeds as follows. First, we compute the neighborhood of $f$. Then, for each point $e_1 \in E_1$, we compute the distance between $e_1$ and the nearest point to $e_1$ in the neighborhood of $f$. We call this distance *search threshold*. Then, we determine the count of the points in the blocks that are *completely* included within the search threshold. If the count exceeds $k_\bowtie$, i.e., the $k$ value of the join, then the neighborhood of $e_1$ cannot intersect the neighborhood of $f$. Thus, it is useless to compute the neighborhood of $e_1$. Otherwise, we compute the neighborhood of $e_1$, intersect it with the neighborhood of $f$, and produce pairs of the form $(e_1, i)$, where $i$ belongs to the intersection. An illustration of the Counting algorithm is given in Figure 4.

---
[1]Notice that the same challenge exists if the selection is a spatial range (e.g., rectangle), or a relational attribute-based selection

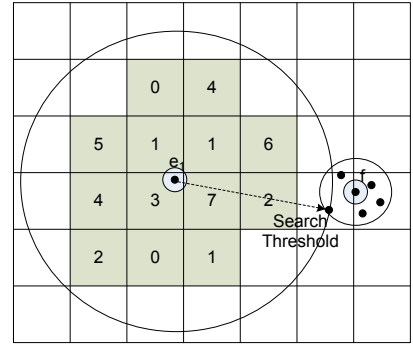

Figure 4: The small circle to the right confines the neighborhood of $f$ in $E_2$. The search threshold is the distance between $e_1$ and the nearest to it in the neighborhood of $f$. If the count of the points of $E_2$ in the gray blocks (i.e., blocks that are *completely* included within the search threshold) exceeds $k_\bowtie$, point $e_1$ **is ignored.**

Procedure 1 gives pseudocode for the algorithm. We assume the existence of Method getkNN(p, k) that returns the neighborhood of a point, say $p$, and Method intersect(P, Q) that returns the set-intersection between two sets of points, say $P$ and $Q$. We use both methods throughout the paper.

---
**Procedure 1** $k$NN-join $k$NN-select (Counting)
---
1: $nbr_f \leftarrow getkNN(f, k_\sigma)$ // Neighborhood of $f$
2: $outputPairs \leftarrow \emptyset$
3: **for** $(e_1 \in E_1)$ **do**
4:     // Get the distance from $e_1$ to the nearest point to it in $nbr_f$
5:     $searchThreshold \leftarrow distance(e_1, nbr_f.nearest)$
6:     $count \leftarrow 0$
7:     $maxOrder \leftarrow$ A MAXDIST ordering of $E_2$ blocks from $e_1$
8:     **while** $count \leq k_\bowtie$ **do**
9:       $block \leftarrow maxOrder.next()$
10:    **if** $MAXDIST(block, e_1) > searchThreshold$ **then**
11:       **break**
12:    **end if**
13:    $count \leftarrow count + block.numberOfPoints$
14:    **end while**
15:    **if** $count \leq k_\bowtie$ **then**
16:       $nbr_{e_1} \leftarrow getkNN(e_1, k_\bowtie)$ // Neighborhood of $e_1$
17:       $intersection \leftarrow intersect(nbr_f, nbr_{e_1})$
18:       **for** $(i \in intersection)$ **do**
19:         $outputPairs.add(e_1, i)$
20:       **end for**
21:    **end if**
22: **end for**
23: **return** $outputPairs$

---

To determine the count of points in the blocks of $E_2$ that are completely included within the search threshold, we scan the blocks of the index of $E_2$ in increasing order of their MAXDIST from $e_1$. We keep accumulating the count of the points in the encountered blocks. As mentioned in Section 2, we assume that the index stores the count of the points in each block. Once a block, say $BM$, having its MAXDIST greater than the search threshold is encountered, we stop (see Line 11). The reason is that $BM$ and the ones to follow are not completely included within the search threshold. Also, we stop if the number of points in the encountered blocks exceeds $k_\bowtie$ (see Line 8). In this case, processing more blocks would result in a count that is also greater than $k_\bowtie$.

## 3.2 Block-Marking Algorithm

The Block-Marking algorithm proceeds as follows. First, we compute the neighborhood of $f$. Then, before performing the join, we perform a preprocessing step for all the blocks of $E_1$. For each block, we determine whether points located inside the block can contribute to the results of the query or not. If it is the case that no point $e_1 \in E_1$ in the block can contribute to the results of the query, we mark the entire block *Non-Contributing*. Otherwise, the block is marked *Contributing*.

After the preprocessing step, we scan the Contributing blocks of $E_1$. Non-Contributing blocks are ignored. For each point $e_1$ in a Contributing block, we compute $e_1$'s neighborhood, intersect it with the neighborhood of $f$, and produce pairs of the form $(e_1, i)$, where $i$ is a point that belong to the intersection. Procedure 2 gives pseudocode for the algorithm. Line 2 calls the preprocessing step through Procedure 3 listed next.

**Procedure 2** $k$NN-join $k$NN-select (Block-Marking)
1: $nbr_f \leftarrow getkNN(f, k_\sigma)$ // Neighborhood of $f$
2: $contriburingBlocks \leftarrow preprocess(nbr_f)$
3: $outputPairs \leftarrow \emptyset$
4: **for** $(block \in contriburingBlocks)$ **do**
5:   **for** $(e_1 \in block)$ **do**
6:     $nbr_{e_1} \leftarrow getkNN(e_1, k_\bowtie)$ // Neighborhood of $e_1$
7:     $intersection \leftarrow intersect(nbr_f, nbr_{e_1})$
8:     **for** $(i \in intersection)$ **do**
9:       $outputPairs.add(e_1, i)$
10:     **end for**
11:   **end for**
12: **end for**
13: **return** $outputPairs$

### 3.2.1 Efficient Preprocessing

To determine whether a block is Contributing or not, we compute the neighborhood of the *center* of the block.[2] Then, the distance between the center and the farthest of its neighbors is determined, and is added to the length of the diagonal of the block forming a search threshold. If no point in neighborhood of $f$ is within the search threshold, then we mark the entire block Non-Contributing. In this case, any point, say $p$, in the block will have $k_\bowtie$ or more points that are nearer to $p$ than any point in the neighborhood of $f$.

An equivalent, yet cheaper check can be described as follows. Refer to Figure 5 for illustration. Consider a block, say $NC$, e.g.,

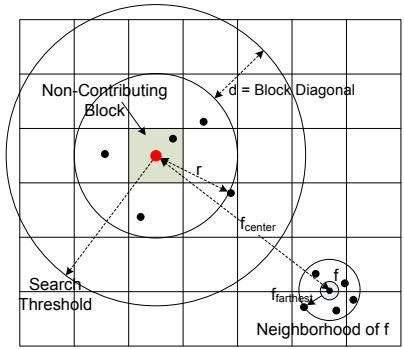

**Figure 5: A block is marked Non-Contributing if** $(r + d + f_{farthest}) < f_{center}$.

---
[2]We discuss the reason behind choosing the center of the block later in this section.

the gray block in Figure 5. Let $r$ be the distance between the center of $NC$ and the farthest of $NC$'s neighbors, $d$ be the length of the diagonal of $NC$, and $f_{farthest}$ be the distance between $f$ and the farthest of $f$'s neighbors, and $f_{center}$ be the distance between $f$ and the center of $NC$. $NC$ is marked Non-Contributing if:

$$(r + d + f_{farthest}) < f_{center}.$$

A brute-force approach for the preprocessing phase is to scan each block in $E_1$, compute the neighborhood of its center, and perform the check described above to determine whether the block is Contributing or not. A more efficient approach is described below.

We scan the blocks of $E_1$ in MINDIST order from $f$. When a block, say $NC$, is marked Non-Contributing, the MAXDIST, say $M$, between $NC$ and $f$ is determined. If all the following encountered blocks are also marked Non-Contributing, then we stop scanning any more blocks when we encounter a block of MINDIST at least $M$. Otherwise, if any of the next encountered blocks is not marked Non-Contributing, then this cycle is repeated. The idea of this approach is to determine a contour (complete cycle) of blocks such that all the blocks in the contour are Non-Contributing. All the blocks outside that contour are considered Non-Contributing without further processing. This is illustrated in Figure 6. Procedure 3 gives pseudocode for the preprocessing phase.

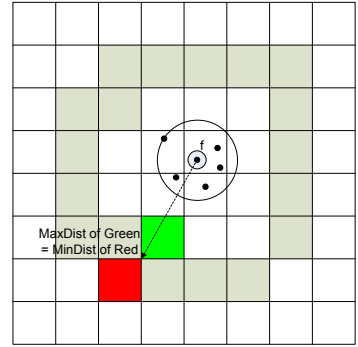

**Figure 6: The preprocessing phase. The green block is a Non-Contributing block. All the next scanned blocks are also Non-Contributing (the contour of gray blocks). Processing stops when the red block is encountered, since its MINDIST from $f$ equals the MAXDIST of the green block from $f$. All the next blocks (outside the gray contour) are considered Non-Contributing without further processing.**

### 3.2.2 Why Choose the Center of the Block?

An important question to address is: If we choose any location, say $c$, other than the center of the block, will this result in a tighter (smaller) search threshold without falsely marking the block Non-Contributing?

THEOREM 1. *The search threshold is minimum if $c$ is the center of the block.*

PROOF. The search threshold is determined by:

1. the distance between $c$ and the farthest of its neighbors, and
2. an added distance, say $x$, that is the length of the diagonal of the block in case $c$ is the center of the block.

The purpose of the added distance $x$ is to cover the neighborhood of any point in the block, i.e., guarantee that the neighborhood of any point in the block does not intersect the neighborhood of $f$.

**Procedure 3** Preprocess Blocks (Block-Marking)

**Terms**: $nbr_f$: The neighborhood of $f$. $M$: MAXDIST between $f$ and the first Non-Contributing block encountered in the cycle (e.g., the green block in the figure).

1: // $f_{farthest}$ is the distance between $f$ and the farthest of its neighbors
2: $f_{farthest} \leftarrow distance(f, nbr_f.farthest)$
3: $contributingBlocks \leftarrow \emptyset$
4: $M \leftarrow 0$
5: $minOrder \leftarrow$ A MINDIST ordering of $E_1$ blocks from $f$
6: **for** $(block \in minOrder)$ **do**
7:   **if** $(block.MINDIST(f) \geq M)$ **then**
8:     **break** // All the remaining blocks are Non-Contributing
9:   **end if**
10:   $nbr \leftarrow getkNN(block.center, k_{\bowtie})$ // Neighborhood of center
11:   // $r$ is the distance between center and the farthest of its neighbors
12:   $r \leftarrow distance(block.center, nbr.farthest)$
13:   $f_{center} \leftarrow distance(block.center, f)$
14:   **if** $(r + block.diagonal + f_{farthest} < f_{center})$ **then**
15:     // Non-Contributing block
16:     **if** $(M = 0)$ **then**
17:       // First Non-Contributing block in the cycle
18:       $M \leftarrow block.MAXDIST(f)$
19:     **end if**
20:   **else**
21:     $contributingBlocks.add(block)$
22:     $M \leftarrow 0$ // Start another cycle
23:   **end if**
24: **end for**
25: **return** $contributingBlocks$

Assume that we randomly select the location of $c$, and compute its neighborhood. Refer to Figure 7 for illustration. The farthest location to $c$ in the block is the top-left corner of the block, say $t$. $\overline{ct} = y$.[3] Point $a$ is the farthest point to $c$ in its neighborhood. $\overline{ac} = r$. Point $b$ is the nearest point to $t$ in the neighborhood of $f$. The region bounded by the search threshold does not intersect the neighborhood of $f$ as shown.

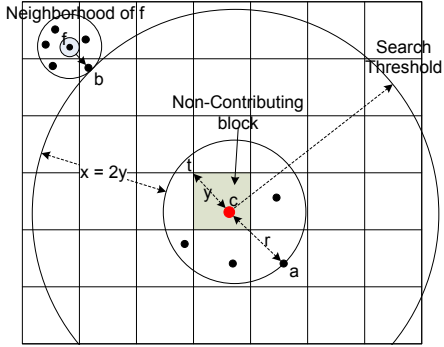

**Figure 7: The effect of choosing any point other than the center of the block to compute the neighborhood for.** $x = 2y$ **is a** *tight* **lower bound for the added distance** $x$ **that guarantees the correct coverage of the search threshold.**

Observe that in Figure 7, we illustrate a bounding case in which the three positions $a$, $b$, and $c$ are collinear and are on the diagonal of the block (or its extension). Point $t$ is in the middle of the distance between Points $a$ and $b$, i.e., $\overline{ta} = \overline{tb} = (y + r)$. Any point inside the block other than $t$ will have distance to Point $a$ that is $< (y + r)$, and also will have distance to point $b$ that is $> (y + r)$.

---
[3]To refer to the distance between two points, say $p_1$ and $p_2$, we use the notation $\overline{p_1p_2}$.

If $x > 2y$ then $\overline{tb} > (y + r)$. For any point inside the block, the distance to Point $a$ will be $< (y + r)$, and the distance to point $b$ will be $> (y + r)$. This means that the neighborhood of any point in the block cannot intersect with the neighborhood of $f$, i.e., the block is correctly marked Non-Contributing.

If $x < 2y$ then $\overline{tb} < (y + r)$. For Point $t$, Point $b$ will be nearer than Point $a$. So, even though no point in the neighborhood of $f$ is within the search threshold, the neighborhood of a point at the top-left corner will intersect the neighborhood of $f$, i.e., the block is falsely marked Non-Contributing.

Thus, $x = 2y$ is a *tight* lower bound for the added distance $x$. And since $y$ is the distance from $c$ to the farthest corner of the block, $y$ is minimum if $c$ is the center of the block. For this reason, the search threshold is minimum if $c$ is the center of the block. □

### 3.3 Counting vs. Block-Marking

An important question to address is: How do we choose between the Counting and Block-Marking algorithms? Observe that the Counting algorithm does not require a preprocessing phase, i.e., once the query is issued, points of the outer relation are processed. However, the Block-Marking algorithm requires a preprocessing phase to determine the Contributing and Non-Contributing blocks. Although this is a winning point for the Counting algorithm, the Block-Marking algorithm always has better opportunities for being faster.

In the Counting algorithm, for every point in the outer relation, the number of points in the blocks that are within the search threshold has to be determined. In other words, the Counting algorithm poses a per-tuple overhead. On the other hand, the Block-Marking algorithm has a per-block overhead (to determine the Contributing blocks). Furthermore, as discussed in Section 3.2.1, this per-block overhead does not affect all the blocks of the outer relation. The reason is that the preprocessing phase stops when a contour of Non-Contributing blocks is encountered.

As we illustrate in Section 6, when the number of points in the outer relation is small, the Counting algorithm has better performance. In this case, because the density of the points is relatively low, the overhead of the preprocessing phase of the Block-Marking algorithm is relatively high as it requires computing the neighborhood of the centers of many blocks without significant payoff. On the other hand, when the number of points in the outer relation is relatively high, i.e., high density, the Block-Marking algorithm has better performance because entire blocks will be excluded from the join. On the contrary, the Counting algorithm will have to process every point.

## 4. TWO KNN-JOINS

As mentioned in Section 1, the $k$NN-join is not a symmetric operation, i.e., the two expressions $(E_1 \bowtie_{kNN} E_2) \cap (E_2 \bowtie_{kNN} E_3)$ and $(E_1 \bowtie_{kNN} E_2) \cap (E_3 \bowtie_{kNN} E_2)$ are not equivalent. We call the joins in the former expression *chained* ($E_1 \rightarrow E_2 \rightarrow E_3$), and the joins in the latter expression *unchained*.

### 4.1 Unchained kNN-Joins

Consider a query on three data sets, say $A$, $B$, and $C$. The query is to retrieve the triplets $(a, b, c)$, where $a \in A$, $b \in B$, and $c \in C$, such that $b$ is a $k_{A-B}$ nearest neighbor of $a$, and $b$ is a $k_{C-B}$ nearest neighbor of $c$. Figures 8 and 9 give two possible QEPs for the query. In both figures, solid lines indicate the $k$NN-join performed first, and dashed lines indicate the $k$NN-join performed at the end.

Although both QEPs seem to be legitimate, they produce different results; surprisingly none of them is correct. The reason is that if either join is performed first, then it filters out the input of the

1104

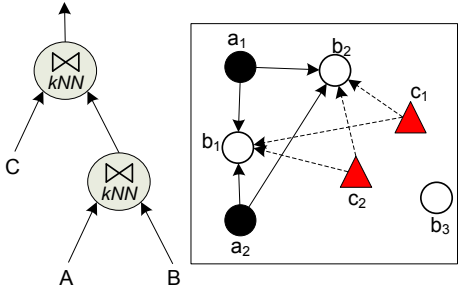

**Figure 8:** $(A \bowtie_{kNN} B)$ **is evaluated before** $(C \bowtie_{kNN} B)$. $k_{A-B} = k_{C-B} = 2$. **The resulting triplets are:** $(a_1, b_1, c_1)$, $(a_1, b_1, c_2)$, $(a_2, b_1, c_1)$, $(a_2, b_1, c_2)$, $(a_1, b_2, c_1)$, $(a_1, b_2, c_2)$, $(a_2, b_2, c_1)$, **and** $(a_2, b_2, c_2)$.

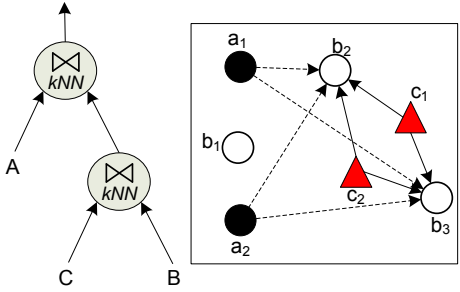

**Figure 9:** $(C \bowtie_{kNN} B)$ **is evaluated before** $(A \bowtie_{kNN} B)$. $k_{A-B} = k_{C-B} = 2$. **The resulting triplets are:** $(a_1, b_3, c_1)$, $(a_1, b_3, c_2)$, $(a_2, b_3, c_1)$, $(a_2, b_3, c_2)$, $(a_1, b_2, c_1)$, $(a_1, b_2, c_2)$, $(a_2, b_2, c_1)$, **and** $(a_2, b_2, c_2)$.

inner relation of the other join. For example, in Figure 8, when $(A \bowtie_{kNN} B)$ is performed first, point $b_3$ is filtered out and will not be in the neighborhood of any point $c \in C$. Similarly, in Figure 9, when $(C \bowtie_{kNN} B)$ is performed first, point $b_1$ is filtered out and will not be in the neighborhood of any point $a \in A$. Each QEP is equivalent to pushing a selection on the inner relation of a $k$NN-join, which has been proven to be invalid earlier in the paper.

According to [19], to evaluate a query with two unchained $k$NN-joins, each join has to be evaluated *independently*. The results of the two joins are *combined* using some operation that has the same flavor as intersection. This operation takes as input the two sets of pairs of the outputs of the two joins, and returns the matching pairs that have the same $B$ component, i.e., intersects the two sets of pairs on $B$, which we denote by $\cap_B$. This is illustrated in the QEP in Figure 10.

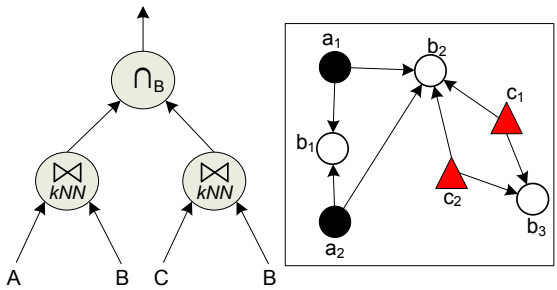

**Figure 10: The two joins** $(C \bowtie_{kNN} B)$ **and** $(A \bowtie_{kNN} B)$ **are evaluated independently.** $k_{A-B} = k_{C-B} = 2$. **The resulting triplets are:** $(a_1, b_2, c_1)$, $(a_1, b_2, c_2)$, $(a_2, b_2, c_1)$, **and** $(a_2, b_2, c_2)$.

### 4.1.1 Efficient Evaluation

Consider the QEP in Figure 10 for evaluating unchained $k$NN-joins. Notice that because the two joins are evaluated independently, we can start with either join. Without loss of generality, assume that the execution starts by evaluating the join $(A \bowtie_{kNN} B)$. We study the issue of choosing the optimal join order later in this section. This QEP is efficient if every point $c \in C$ is part of the final results of the query. As we show next, if some points in $C$ do not contribute to the results of the query, computing their neighborhood is redundant, and can be avoided without losing the correctness of evaluation. This is illustrated in Figure 11 that shows the distribution of the data sets $A$, $B$, and $C$.

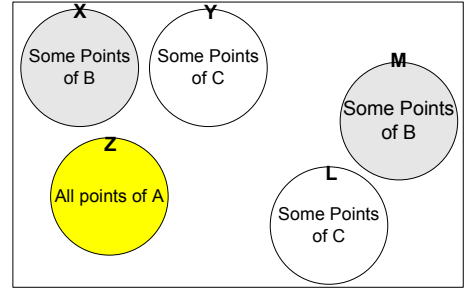

**Figure 11: For points in Circle $L$, the join** $(C \bowtie_{kNN} B)$ **is redundant and its computation can be avoided.**

In Figure 11, points of Set $A$ are in Circle $Z$, points of Set $B$ are divided between Circles $X$ and $M$, and points of Set $C$ are divided between Circles $Y$ and $L$. The points in Circle $M$ confine the neighborhood of the points in Circle $L$. The points in Circle $X$ confine the neighborhood of the points in Circle $Y$. The points in Circle $X$ confine also the neighborhood of points in Circle $Z$. For all the points in Circle $L$, performing the join $(C \bowtie_{kNN} B)$ is redundant because its result will never intersect the result of the join $(A \bowtie_{kNN} B)$ as the join result of the latter is fully contained in Circle $X$. On the other hand, for the points in Circle $Y$, performing the join $(C \bowtie_{kNN} B)$ is essential, because its result will be in Circle $X$ that also contains the result of the join $(A \bowtie_{kNN} B)$.

To efficiently evaluate a query with two unchained $k$NN-joins $(A \bowtie_{kNN} B)$ and $(C \bowtie_{kNN} B)$, we follow the following procedure. After evaluating the join $(A \bowtie_{kNN} B)$, we determine the blocks of $B$ that contain points $b \in B$ that belong to the resulting pairs $(a, b)$, where $a \in A$. We mark these blocks as *Candidate* blocks. All the other blocks are marked as *Safe* blocks. For example, in Figure 11, Circle $X$ is a Candidate block, and Circle $M$ is a Safe block.

Before evaluating the join $(C \bowtie_{kNN} B)$, we do a preprocessing step similar to the preprocessing step of the Block-Marking technique in Section 3.2. In this preprocessing step, we scan all the blocks of $C$ to determine the blocks that are contributing or non-contributing to the results of the query. For each block, we compute the neighborhood of its center. Then, the distance from the center to the farthest point in its neighbors is determined, and is added to the length of the diagonal of the block to form a search threshold as in Figure 12. We mark the block Non-Contributing if all the blocks that are fully or partially contained within the search threshold are Safe.

After the preprocessing step, we scan the Contributing blocks of $C$. Non-Contributing blocks are ignored. For each point, say $c$,

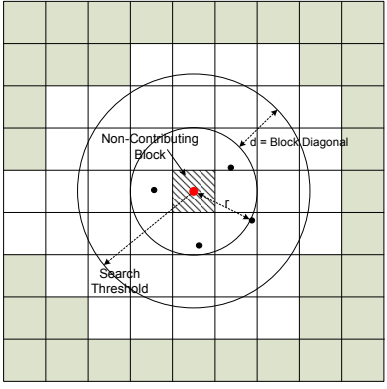

**Figure 12: gray blocks are Candidate blocks. White blocks are Safe blocks. A block is Non-Contributing if the blocks that are fully or partially contained within its search threshold are Safe.**

in a Contributing block, we compute $c$'s neighborhood, and produce pairs of the form $(c, b)$ that we intersect on $B$ (i.e., $\cap_B$) with the computed pairs of the join $(A \bowtie_{kNN} B)$. Procedure 4 gives pseudocode for the algorithm.

---
**Procedure 4** Unchained $k$NN-joins (Block-Marking)

**Terms**: $A, B, C$: The input relations of the two joins. $k_{A-B}, k_{C-B}$: The $k$ values of the joins $(A \bowtie_{kNN} B)$ and $(C \bowtie_{kNN} B)$, respectively.
1: // Perform the join $(A \bowtie_{kNN} B)$
2: $ABpairs \leftarrow kNNJoin(A, B, k_{A-B})$
3: $BPointsInAB \leftarrow project(ABpairs)$ // Project on $B$
4: // Determine Candidate blocks of $C$ (a block is Safe by default)
5: **for** $(b \in BPointsInAB)$ **do**
6:    $block \leftarrow C.index.locate(b)$
7:    $block.isSafe \leftarrow false$
8: **end for**
9: // Preprocess the blocks of $C$ to determine the Contributing ones
10: $contributingBlocks \leftarrow \emptyset$
11: **for** $(block \in C.index)$ **do**
12:    **if** $(block.isSafe = false)$ **then**
13:      $contributingBlocks.add(block)$
14:    **else**
15:      $nbr \leftarrow getkNN(block.center, k_{C-B})$
16:      $r \leftarrow distance(block.center, nbr.farthest)$
17:      $searchThreshold \leftarrow r + block.diagonal$
18:      **if** (any block within $searchThreshold$ is Candidate) **then**
19:         $contributingBlocks.add(block)$
20:      **end if**
21:    **end if**
22: **end for**
23: // Perform the join $(C \bowtie_{kNN} B)$ and intersect on $B$
24: $outputTriplets \leftarrow \emptyset$
25: **for** $(block \in contriburingBlocks)$ **do**
26:    **for** $(c \in block)$ **do**
27:      $nbr_c \leftarrow getkNN(c, k_{C-B})$ // Neighborhood of $c$
28:      **for** $((a, b) \in ABPairs)$ **do**
29:         **if** $(b \in nbr_c)$ **then**
30:            $outputTriplets.add(a, b, c)$
31:         **end if**
32:      **end for**
33:    **end for**
34: **end for**
35: **return** $outputTriplets$

---

A simple optimization for the preprocessing phase is to process only the Safe blocks. This is because a Candidate block is never marked Non-Contributing as its center is not contained in a Safe block (refer to the check in Line 12 of Procedure 4).

### 4.1.2 Join Order

In the QEP of Figure 10, each $k$NN-join is evaluated independently. Thus, changing the order of the two unchained $k$NN-joins leads to the same results for the query. However, choosing which join to evaluate first can affect the number of Candidate and Safe blocks, and hence directly impacts the number of Non-Contributing (pruned) blocks in the second join. Hence, the question: Which of the joins $(A \bowtie_{kNN} B)$ and $(C \bowtie_{kNN} B)$ should be evaluated first?

Consider the case when the points in $A$ and $B$ are uniformly distributed and cover the whole space, while the points in $C$ are clustered inside a certain region, say $R$. If we perform the join $(A \bowtie_{kNN} B)$ first, there will be no Safe blocks because the neighborhood of the points of $A$ will cover all the blocks in $B$ due to the uniformity in data distribution. This means that all the blocks of $C$ will be Contributing, i.e., no pruning will take place. On the other hand, if we perform the join $(C \bowtie_{kNN} B)$ first, the Candidate blocks will be only in Region $R$ and its surroundings. This means that there will be several Safe blocks. This will result in Non-Contributing blocks in $A$ that are pruned during the other join $(A \bowtie_{kNN} B)$.

In conclusion, considering $A$ and $C$ as the outer relations of two unchained $k$NN-joins:

- If either $A$ or $C$ is clustered, the evaluation of the query should start with the join of the clustered relation. As a consequence, blocks of the inner relation (e.g., $B$) will have higher chance to be Safe. This would maximize the number of Non-Contributing blocks in the outer relation of the second join, and hence these blocks will be pruned.

- If both $A$ and $C$ are clustered, the evaluation of the query should start with the join of the relation that has less cluster coverage, i.e., the relation with clusters that cover smaller area. This increases the chance of pruning in the second join.

- If both $A$ and $C$ are uniformly distributed, it is better to use the QEP of Figure 10, i.e., perform both joins independently. If Procedure 4 is applied, then there will be a preprocessing overhead (to mark the blocks) without payoff. The reason is that all the blocks of the outer relation of the second join will be Contributing, i.e., no pruning will occur.

In Section 6.2.1, we exploit various data distributions and cluster setups that demonstrate the effects depicted in the above cases.

## 4.2 Chained kNN-Joins

Consider a query on three data sets, say $A$, $B$, and $C$. The query is to retrieve the triplets $(a, b, c)$, where $a \in A$, $b \in B$, and $c \in C$, such that $b$ is a $k_{A-B}$ nearest neighbor of $a$, and $c$ is a $k_{C-B}$ nearest neighbor of $b$. The query can be evaluated in a variety of ways as Figure 13 illustrates. The three QEPs in the figure produce the same results for the query, i.e., the following relation holds [19]:

$$(A \bowtie_{kNN} B) \cap (B \bowtie_{kNN} C) \equiv \\ (A \bowtie_{kNN} B) \bowtie_{kNN} C \equiv \\ A \bowtie_{kNN} (B \bowtie_{kNN} C).$$

The correctness of the above relation can be explained as follows. The join $(A \bowtie_{kNN} B)$ can be viewed as a selection on the outer relation of the join $(B \bowtie_{kNN} C)$ (i.e., selection on $B$). Similar to the discussions in Section 3, pushing a selection on the outer relation of a $k$NN-join does not affect the correctness of evaluation. That is why performing the join $(A \bowtie_{kNN} B)$ before or after the join $(B \bowtie_{kNN} C)$ leads to the same results.

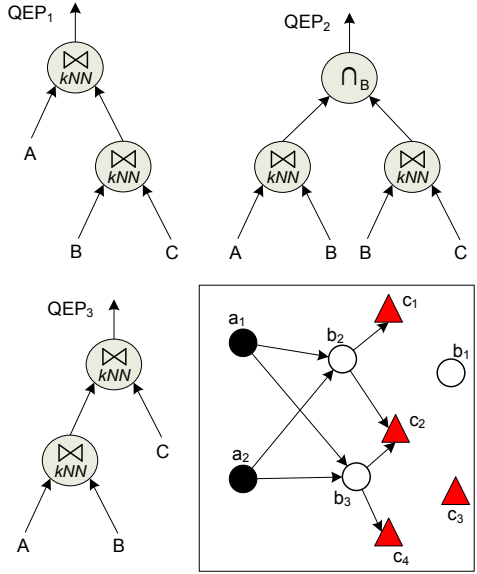

**Figure 13:** $k_{A-B} = k_{B-C} = 2$. **Three different QEPs for a query with two chained $k$NN-joins. The three QEPs result in the same triplets:** $(a_1, b_2, c_1)$, $(a_1, b_2, c_2)$, $(a_2, b_2, c_1)$, $(a_2, b_2, c_2)$, $(a_1, b_3, c_2)$, $(a_1, b_3, c_4)$, $(a_2, b_3, c_2)$, **and** $(a_2, b_3, c_4)$.

### 4.2.1 Efficient Evaluation

Although the three QEPs in Figure 13 produce the same results, they have different performance. The following points illustrate the pros and cons of each QEP.

- QEP$_1$ is a right deep plan; the results of the join $(B \bowtie_{kNN} C)$ have to be materialized before proceeding with the other join. This is a major drawback, because no output can be produced until after the join $(B \bowtie_{kNN} C)$ is complete. Moreover, performing the join $(B \bowtie_{kNN} C)$ first implies that some redundant computations will be performed, e.g., getting the neighborhood of $b_1$ although it will never appear in the results of the query as it is not in the neighborhood of any point $a \in A$.

- QEP$_2$ has an extra operator; $\cap_B$; to intersect the results of both joins on $B$. Moreover, QEP$_2$ suffers the same redundant computations as QEP$_1$, since QEP$_2$ blindly computes the neighborhood of every point $b \in B$ regardless of whether or not $b$ appears in the results of the query.

- QEP$_3$ avoids the redundant computations of QEP$_1$ and QEP$_2$. The neighborhood of a point $b \in B$ is computed only if $b$ is produced as a nearest neighbor to a point $a \in A$. Thus, computing the neighborhood of $b_1$ is avoided in this QEP. This results in remarkable performance gains for QEP$_3$ in comparison to QEP$_1$ and QEP$_2$ especially for relations that have clusters of points. Clusters of points in $B$ that are not in the neighborhood of any point $a \in A$ are pruned in the joins of QEP$_3$. However, both QEP$_1$ and QEP$_2$ will have to process all the clusters. On the other hand, QEP$_3$ suffers some repeated computations. In particular, this happens for every point $b$ that is in the neighborhood of more than one point in $A$. For example, computing the neighborhood of $b_2$ is performed twice because $b_2$ appears in the neighborhood of both $a_1$ and $a_2$. Similarly, the neighborhood of $b_3$ is computed twice.

To avoid the repeated computations in QEP$_3$, we *cache* the results of the join $(B \bowtie_{kNN} C)$ in a hash table, where $b \in B$ is the key, and the value is the neighborhood of $b$. Whenever a pair $(a, b)$ is produced from the join $(A \bowtie_{kNN} B)$, the hash table is probed to check if an entry corresponding to $b$ exists. If such entry exists, the neighborhood of $b$ is retrieved from the hash table. Otherwise, the neighborhood of $b$ is computed. As we show in Section 6, caching the results of the join $(B \bowtie_{kNN} C)$ significantly improves the performance of QEP$_3$, and thus outperforms both QEP$_1$ and QEP$_2$.

## 5. TWO KNN-SELECTS

### 5.1 Correct Conceptual Evaluation

When two $k$NN-select predicates are combined in a single query, different QEPs that seem to be legitimate can produce different results. The following example illustrates such ambiguity in the evaluation of a query with two $k$NN-selects.

Assume that a person gets a new job in a city different from where he lives. He decides to move with his family to the new city, and considers buying a new house such that the new house is close to both his work and the school of his children. He wants to select candidate houses to choose from such that these houses are among the closest five houses to both his work and the school.

Figures 14 and 15 give two different QEPs for the above query with the corresponding resulting houses. In both figures, solid lines indicate the $k$NN-select predicate performed first, and dashed lines indicate the $k$NN-select predicate performed second.

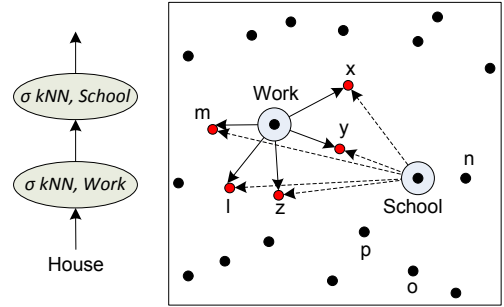

**Figure 14:** A QEP with $\sigma_{kNN,Work}(House)$ **performed before** $\sigma_{kNN,School}(House)$. **The resulting houses are:** $x$, $y$, $l$, $m$, **and** $z$.

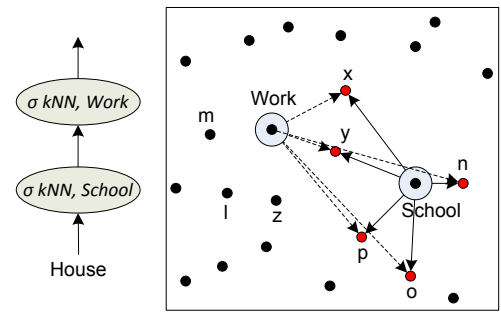

**Figure 15:** A QEP with $\sigma_{kNN,School}(House)$ **performed before** $\sigma_{kNN,Work}(House)$. **The resulting houses are:** $x$, $y$, $n$, $p$, **and** $o$.

Although the QEPs in Figures 14 and 15 seem legitimate, they produce different results. Surprisingly, both results are wrong. The

1107

reason is that when any of the two *k*NN selects is performed first, it filters out the input of the other *k*NN select. The scope of the *k*NN select performed at the end will be limited to only the *k* points that qualify the first *k*NN select. For example, in Figure 14, $\sigma_{kNN,School}(House)$ has nothing to select from except the five houses that $\sigma_{kNN,Work}(House)$ returns. Similarly, in Figure 15, $\sigma_{kNN,Work}(House)$ has nothing to select from except the five houses that $\sigma_{kNN,School}(House)$ returns.

According to [19], for the above query to be correctly evaluated, each *k*NN-select predicate has to be evaluated *independently*. Then, the results of applying both predicates are intersected. This is illustrated in the QEP in Figure 16.

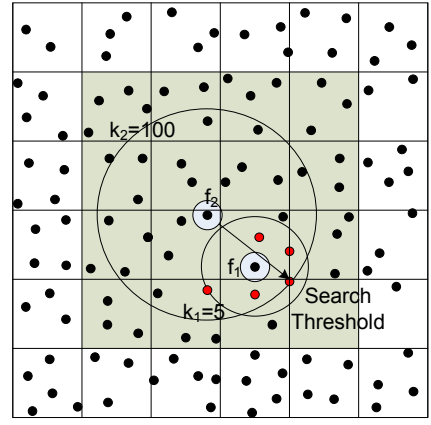

Figure 17: The *search threshold* is the distance between $f_2$ and the farthest to it in the neighborhood of $f_1$. The gray blocks represent the locality of $f_2$. A block is added to the locality of $f_2$ if its MINDIST from $f_2$ is less than the search threshold.

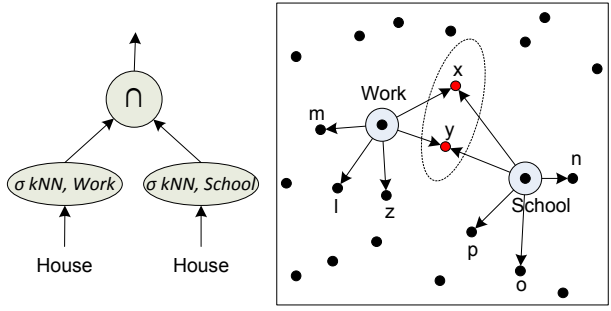

Figure 16: The correct QEP for a query with two *k*NN-select predicates. Each predicate is evaluated independently, and the results are intersected. The resulting houses are: $x$ and $y$.

## 5.2 Efficient Evaluation

The QEP in Figure 16 for evaluating a query with two *k*NN-selects, say $\sigma_{k_1,f_1}(E)$ and $\sigma_{k_2,f_2}(E)$, is efficient if $k_1 = k_2$. If $k_1 \neq k_2$, the above QEP suffers some redundancy as the following discussions demonstrate.

Consider a query that has the two *k*NN-selects $\sigma_{5,f_1}(E)$ and $\sigma_{100,f_2}(E)$, i.e., $k_1 = 5$ and $k_2 = 100$ (i.e., $k_2$ has a value that is significantly greater than $k_1$). As mentioned in Section 2, in order to compute the neighborhood of a point, say $p$, the locality is first determined. Then, the points inside the blocks of the locality are processed in order to get the closest $k$ points to $p$. The standard approach (as in [15]) of computing the locality is to keep adding blocks to the locality until a total of $k$ points is reached in the encountered blocks. If this approach is applied to $\sigma_{100,f_2}(E)$, the locality of $f_2$ will be large and will cover almost the entire space in which the points reside. In other words, almost all the blocks will be in the locality of $f_2$, and will have to be processed in order to find the neighborhood of $f_2$.

The above approach for computing the locality of $f_2$ is not efficient because it does not consider the neighborhood of $f_1$. In particular, the number of blocks in the locality of $f_2$ can be smaller and still produce correct results. This can be achieved by observing that the neighborhood of $f_1$ is completely included inside the locality of $f_2$.

Because the final result of the query is determined by intersecting the neighborhoods of $f_1$ and $f_2$, this final result cannot include points other than the neighborhood of $f_1$. Consequently, once the neighborhood of $f_1$ is determined, the locality of $f_2$ can be adjusted to cover just the neighborhood of $f_1$. We define the *search threshold* as the distance between $f_2$ and the farthest to it in the neighborhood of $f_1$. A block, say $b$, is added to the locality of $f_2$ only if the MINDIST between $b$ and $f_2$ is less than or equal to the search threshold. Refer to Figure 17 for illustration. This guarantees that the neighborhood of $f_1$ is included in the locality of $f_2$ and in turn, the final result of the query.

---

**Procedure 5** 2-*k*NN-select

**Terms**: $nbr_1, nbr_2$: The neighborhoods of $f_1$ and $f_2$, respectively.
1: **if** $k_1 > k_2$ **then**
2:     swap($k_1, k_2$)
3:     swap($f_1, f_2$)
4: **end if**
5: $nbr_1 \leftarrow getkNN(f_1, k_1)$
6: $searchThreshold \leftarrow \text{distance}(f_2, nbr_1.farthestTof_2)$
7: $f_2.locality \leftarrow \emptyset$
8: $count \leftarrow 0$
9: $maxDistSoFar \leftarrow 0$
10: // Process the blocks in MAXDIST order from $f_2$
11: **while** $count < k_2$ **do**
12:     $block \leftarrow maxOrder.next()$
13:     $count \leftarrow count + block.numberOfPoints$
14:     $maxDistSoFar \leftarrow MAXDIST(block, f_2)$
15:     **if** $MINDIST(block, f_2) \leq searchThreshold$ **then**
16:         $f_2.locality.add(block)$
17:     **end if**
18: **end while**
19: // Process the remaining blocks in MINDIST order from $f_2$
20: **for** ($block \in minOrder$) **do**
21:     **if** $MAXDIST(block, f_2) \leq maxDistSoFar$ **then**
22:         **if** $MINDIST(block, f_2) \leq searchThreshold$ **then**
23:             $f_2.locality.add(block)$
24:         **else**
25:             break
26:         **end if**
27:     **else**
28:         break
29:     **end if**
30: **end for**
31: // Determine the neighborhood of $f_2$ from its locality
32: $nbr_2 \leftarrow getNeighborhood(f_2, f_2.locality)$
33: **return** $intersect(nbr_1, nbr_2)$

---

Procedure 5 gives pseudocode for evaluating two *k*NN-select predicates. The procedure starts by computing the neighborhood of $f_1$, i.e., evaluating the predicate with smaller $k$. To compute the neighborhood of $f_2$, $f_2$'s locality is determined using a slightly different version of the algorithm in [15]. In [15], to determine the locality of a point, say $p$, the blocks of the index are processed in increasing order of their MAXDIST from $p$, and are added to the locality. The counts of the number of points in the blocks are summed up until the total number of points in the encountered blocks exceeds $k$. At this moment, the current value of the MAXDIST, say $M$, is recorded. Afterwards, the remaining blocks are processed in increasing order of their MINDIST from $p$, and are added to the locality until a block the MINDIST of which exceeds $M$ is encountered. All the remaining blocks need not be examined. This procedure for building the locality is proven to guarantee the optimal (minimum) possible number of blocks [15]. We follow the same procedure for computing the locality of $f_2$ except that a block, say $b$, is added to the locality of $f_2$ only if the MINDIST between $b$ and $f_2$ is less than or equal to the search threshold. Refer to Lines 15 and 22, and 25 of Procedure 5. Notice that in Line 25, scanning the blocks in MINDIST order stops when a block of MINDIST greater than the search threshold is encountered.

## 6. EXPERIMENTAL RESULTS

In this section, we study the performance of the proposed optimization techniques. We measure the query execution time. To compute the neighborhood of a point, we implement the locality algorithm as in [15]. All implementations are in Java. Experiments are conducted on a machine running Windows 7 with Intel Core2 Duo CPU at 2.1 GHz and 4 GB of main memory.

Our datasets are mainly generated using BerlinMOD [3]; a benchmark for spatio-temporal database management systems. The data is downloadable through the BerlinMOD website [1] with scale-factor 1.0. In BerlinMOD, about two thousand cars report their movement over Berlin City for 28 days. We remove the time dimension from the data to deal with snapshots of points. Depending on the kind of experiment, we vary the number of points in the datasets, from 32,000 to 2,560,000 data points. A sample snapshot of the data is given in Figure 18. In addition to the BerlinMOD data, and in order to demonstrate some specific effects, we generate our own synthetic data. In particular, for some experiments, we generate clustered data and vary the number of clusters.

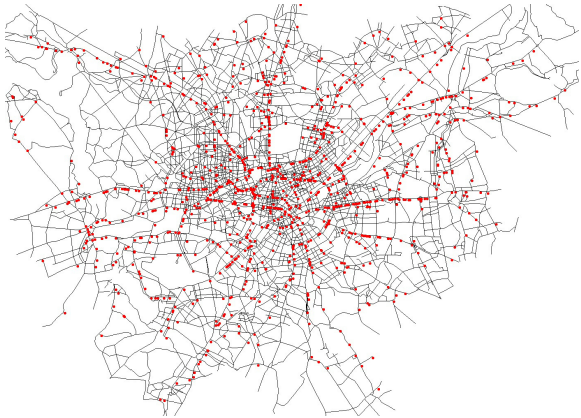

**Figure 18: A sample snapshot of BerlinMOD data plotted on the map of Berlin City.**

We index the data points into a simple grid. Since our algorithms are independent of a specific indexing structure, we choose a grid in order to be able to see the effectiveness of our algorithms even with simple structures. We expect our algorithms to maintain the same effectiveness (if not better) with more robust index implementations, e.g., using variants of the R-tree or the quadtree.

### 6.1 kNN-Select with kNN-Join

In the following experiments, we study the performance of the two proposed algorithms, Counting and Block-Marking, for a query with a $k$NN-select on the inner relation of a $k$NN-join. Figure 19 illustrates that the Block-Marking algorithm outperforms the conceptually correct QEP by orders of magnitude. Blocks of points of the outer relation that do not contribute to the results of the join are detected and are excluded from the join operation. From the figure, increasing the number of points in the outer relation emphasizes the pruning effects of the algorithm.

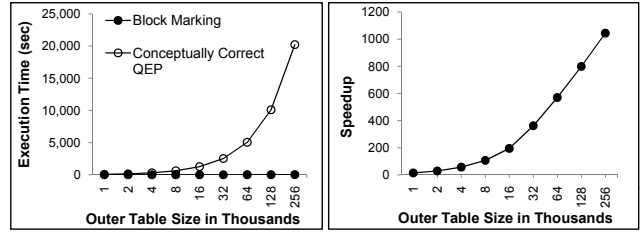

**Figure 19: Execution time of a query with a $k$NN-select on the inner relation of a $k$NN-join. The Block-Marking algorithm outperforms the conceptually correct evaluation plan by three orders of magnitude.**

Figures 20 and 21 compare the performance of the Counting and Block-Marking algorithms. In Figure 20, the number of points in the outer relation is lower than those in Figure 21. As the figures demonstrate, when the number of points in the outer relation is small, the Counting algorithm has better performance. In this case, the density of the points is relatively low, and the overhead of the preprocessing phase required by the Block-Marking algorithm is relatively high because it requires computing the neighborhood of the centers of many blocks without much payoff. On the other hand, when the number of points in the outer relation is high, i.e., the outer relation has high density, the Block-Marking algorithm has better performance because entire blocks are excluded from the join. On the contrary, the Counting algorithm processes every point.

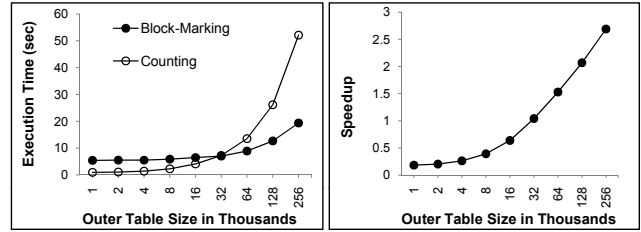

**Figure 20: Execution time of a query with a $k$NN-select on the inner relation of a $k$NN-join. The Counting algorithm has better performance than the Block-Marking algorithm when the number of points in the outer relation is low, and vice versa.**

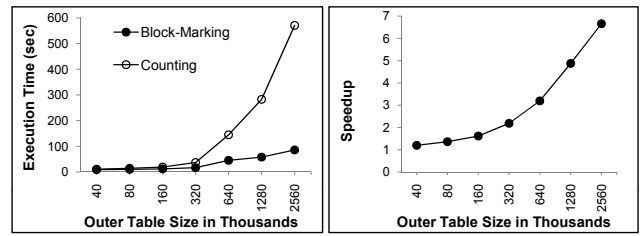

**Figure 21: Execution time of a query with a $k$NN-select on the inner relation of a $k$NN-join. The Block-Marking algorithm has much better performance than the Block-Marking algorithm when the number of points in the outer relation is high.**

## 6.2 Two kNN-Joins

### 6.2.1 Unchained kNN-Joins

In the following experiments, we study the performance of the Block-Marking algorithm for a query with two unchained $k$NN-joins, e.g., $(A \bowtie_{kNN} B)$ and $(C \bowtie_{kNN} B)$. As mentioned in Section 4.1.2, if both $A$ and $C$ are uniformly distributed, then it is better to use the conceptually correct QEP of Figure 10, i.e., perform both joins independently, than to use the Block-Marking algorithm. In that case, if the Block-Marking algorithm is applied, then there will be a preprocessing overhead without payoff.

To demonstrate the pruning effects of the Block-Marking algorithm, we have the following experimental setup. Points of $B$ and $C$ are generated using BerlinMOD. Points of $A$ are generated such that they are clustered inside a certain region. We fix the number of points in $A$ and $B$, and vary the number of points in $C$. Figure 11 illustrates that the Block-Marking algorithm can outper-form the conceptually correct QEP by an order of magnitude. As the figure demonstrates, the Block-Marking algorithm almost has constant performance because it detects the blocks of $C$ that do not contribute to the results of the query, and excludes them from the join $(C \bowtie_{kNN} B)$. However, the conceptually correct QEP has to perform the join for all the points in $C$ regardless of the layout of the data.

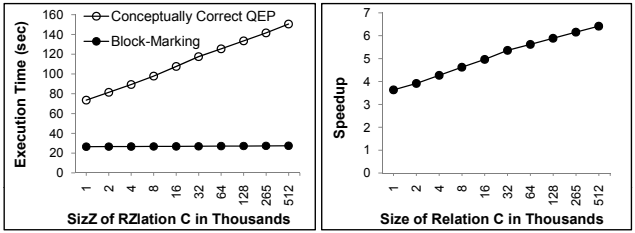

Figure 22: Execution time of a query with two unchained $k$NN-joins $(A \bowtie_{kNN} B)$ and $(C \bowtie_{kNN} B)$. $B$ and $C$ **are uniformly distributed, and** $A$ **is clustered. The Block-Marking algorithm outperforms the conceptually correct QEP by an order of magnitude.**

If both $A$ and $B$ are clustered, then applying the Block-Marking technique can also result in good performance gains. In this case, the evaluation of the query should start with the join of the relation that has less cluster coverage, i.e., the relation the clusters of which cover smaller area. This gives a higher chance for pruning effects in the second join.

To demonstrate this effect, we have the following experimental setup. Points of $B$ are generated using BerlinMOD. We generate clusters of points in $A$ and $C$. All the clusters have the same number of points (4000), have the same area, and are non-overlapping. We vary the number of clusters such that the number of clusters in $A$ is greater than the number of clusters in $C$ by 1, 2, ..., 10. Figure 23 illustrates that starting the evaluation with $(C \bowtie_{kNN} B)$ results in better performance than starting with $(A \bowtie_{kNN} B)$. If the evaluation starts with $(C \bowtie_{kNN} B)$, the Block-Marking algorithm detects the clusters of points in $A$ that do not contribute to the results of the query and excludes them from the join $(A \bowtie_{kNN} B)$. However, starting with $(A \bowtie_{kNN} B)$ will fully compute the join for all the clusters without exclusion.

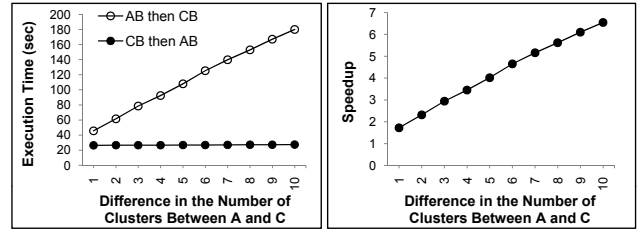

Figure 23: Execution time of a query with two unchained $k$NN-joins $(A \bowtie_{kNN} B)$ **and** $(C \bowtie_{kNN} B)$. $A$ **and** $C$ **are clustered. Varying the difference between the number of clusters in** $A$ **and** $C$**; when the number of clusters in** $C$ **is smaller, starting with** $(C \bowtie_{kNN} B)$ **results in better performance.**

### 6.2.2 Chained kNN-Joins

In the following experiments, we study the performance of the three QEPs of Figure 13, for a query with two chained $k$NN-joins, e.g., $(A \bowtie_{kNN} B)$ and $(B \bowtie_{kNN} C)$. For illustration, we call QEP$_3$: *Nested Join*, and QEP$_2$: *Join Intersection*.

As discussed in Section 4.2, there are two versions of the Nested Join QEP; one that caches the results of the join $(B \bowtie_{kNN} C)$ in a hash table to avoid repeating join computations, and another version that does not do any caching. Figure 24 illustrates that caching the results of the join $(B \bowtie_{kNN} C)$ significantly enhances the performance.

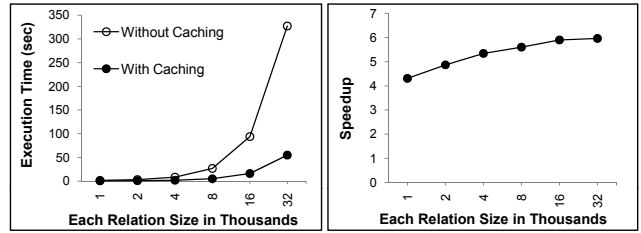

Figure 24: Execution time of a query with two chained $k$NN-joins $(A \bowtie_{kNN} B)$ **and** $(B \bowtie_{kNN} C)$. **Caching the results of the join** $(B \bowtie_{kNN} C)$ **significantly enhances the performance.**

As discussed in Section 4.2, the Join Intersection QEP performs the two joins $(A \bowtie_{kNN} B)$ and $(B \bowtie_{kNN} C)$ independently, and then intersects their results on $B$ (i.e., $\cap_B$. However, the Nested Join QEP performs the join $(B \bowtie_{kNN} C)$ only for points $b \in B$ that are in the neighborhood of one or more points in $A$. When comparing the two QEPs, we find that both plans have almost the same performance if the data points are uniformly distributed. However, as Figure 25 demonstrates, for clustered data, the Nested Join QEP has better performance. We use the version of the Nested Join QEP that caches the results of the join $(C \bowtie_{kNN} B)$. As the number of clusters in $B$ increases, the Nested Join QEP outperforms the Join Intersection QEP. This is because the Join Intersection QEP blindly does both joins without any kind of pruning. However, clusters of points in $B$ that are not in the neighborhood of any point in $A$ are pruned by the Nested Join QEP.

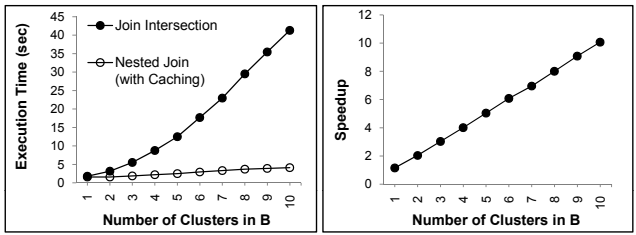

**Figure 25: Execution time of a query with two chained $k$NN-joins $(A \bowtie_{kNN} B)$ and $(B \bowtie_{kNN} C)$. Performance when varying the number of clusters in $B$.**

## 6.3 Two kNN-Selects

In the following experiment, we study the performance of the 2-$k$NN-select algorithm, for a query with two $k$NN-select predicates, e.g., $\sigma_{k_1,f_1}(E)$ and $\sigma_{k_2,f_2}(E)$. Unlike the 2-$k$NN-select algorithm, the conceptually correct QEP fully computes the two $k$NN-selects and then intersects the results, i.e., does not leverage the effect of doing one select and using its output to prune some of the work of the other. In particular, this effect is leveraged by the 2-$k$NN-select algorithm when $k_1$ and $k_2$ have different values.

Figure 26 illustrates how the 2-$k$NN algorithm can outperform the conceptually correct QEP by almost two orders of magnitude. In this experiment, we fix $k_1 = 10$ and vary $k_2$. The x-axis of the figure is $\log_2(k_2/k_1)$. As the ratio $k_1/k_2$ increases, the performance of the conceptually correct QEP degrades. The 2-$k$NN-select algorithm has almost constant performance, as it adjusts the search threshold corresponding to the predicate of higher $k$ value to cover just the output of the predicate of lower $k$ value.

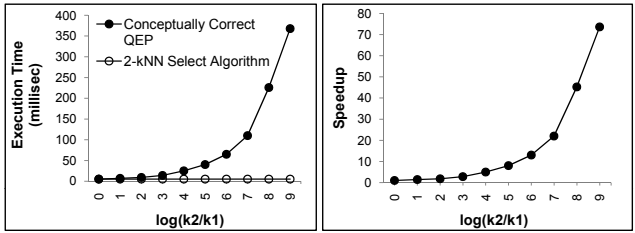

**Figure 26: Execution time of a query with two $k$NN-selects. The 2-$k$NN-select algorithm outperforms the conceptually correct QEP by almost two orders of magnitude.**

## 7. CONCLUSIONS

In this paper, we presented the first complete study for the optimization of queries with two $k$NN predicates. We demonstrated how traditional optimization techniques can compromise the correctness of evaluation for a query that involves two interacting $k$NN predicates. For different combinations of two $k$NN predicates, we presented efficient algorithms that guarantee the correctness of evaluation, and outperform the corresponding conceptually correct QEPs by orders of magnitude.

The algorithms presented in this paper are designed for snapshot queries. Applying further optimization techniques that can support incremental evaluation of continuous queries with two $k$NN predicates is a potential future work. Moreover, we believe that the ideas presented in this paper pave the way towards a query optimizer that can support spatial queries with more than two $k$NN predicates.

## 8. REFERENCES

[1] Secondo Berlinmod. http://dna.fernuni-hagen.de/secondo/BerlinMOD/BerlinMOD.html, 2011.
[2] N. Beckmann, H.-P. Kriegel, R. Schneider, and B. Seeger. The R*-tree: An efficient and robust access method for points and rectangles. In *SIGMOD Conference*, pages 322–331, 1990.
[3] C. Düntgen, T. Behr, and R. H. Güting. Berlinmod: a benchmark for moving object databases. *VLDB Journal*, 18(6):1335–1368, 2009.
[4] H. Garcia-Molina, J. D. Ullman, and J. Widom. *Database System Implementation*. Prentice-Hall, 2000.
[5] R. H. Güting, T. Behr, and J. Xu. Efficient $k$-nearest neighbor search on moving object trajectories. *VLDB Journal*, 19(5):687–714, 2010.
[6] A. Guttman. R-trees: A dynamic index structure for spatial searching. In *SIGMOD Conference*, pages 47–57, 1984.
[7] H.-P. Kriegel, P. Kunath, and M. Renz. R*-tree. In *Encyclopedia of GIS*, pages 987–992. 2008.
[8] M. F. Mokbel and W. G. Aref. Place: A scalable location-aware database server for spatio-temporal data streams. *IEEE Data Eng. Bull.*, 28(3):3–10, 2005.
[9] M. F. Mokbel, X. Xiong, and W. G. Aref. SINA: Scalable incremental processing of continuous queries in spatio-temporal databases. In *SIGMOD Conference*, pages 623–634, 2004.
[10] K. Mouratidis and D. Papadias. Continuous nearest neighbor queries over sliding windows. *IEEE Trans. Knowl. Data Eng.*, 19(6):789–803, 2007.
[11] K. Mouratidis, M. L. Yiu, D. Papadias, and N. Mamoulis. Continuous nearest neighbor monitoring in road networks. In *VLDB*, pages 43–54, 2006.
[12] S. Nutanong, R. Zhang, E. Tanin, and L. Kulik. Analysis and evaluation of V*-$k$NN: an efficient algorithm for moving $k$NN queries. *VLDB Journal*, 19(3):307–332, 2010.
[13] N. Roussopoulos, S. Kelley, and F. Vincent. Nearest neighbor queries. In *SIGMOD Conference*, pages 71–79, 1995.
[14] H. Samet. *Foundations of Multidimensional and Metric Data Structures*. Morgan Kaufmann Publishers Inc., 2006.
[15] J. Sankaranarayanan, H. Samet, and A. Varshney. A fast all nearest neighbor algorithm for applications involving large point-clouds. *Computers & Graphics*, 31(2):157–174, 2007.
[16] Y. N. Silva, A. M. Aly, W. G. Aref, and P.-Å. Larson. Simdb: a similarity-aware database system. In *SIGMOD Conference*, pages 1243–1246, 2010.
[17] Y. N. Silva, W. G. Aref, and M. H. Ali. Similarity group-by. In *ICDE*, pages 904–915, 2009.
[18] Y. N. Silva, W. G. Aref, and M. H. Ali. The similarity join database operator. In *ICDE*, pages 892–903, 2010.
[19] Y. N. Silva, W. G. Aref, P.-Å. Larson, and M. H. Ali. Similarity-aware query processing and optimization. Technical Report 12-006, Department of Computer Science, Purdue University, 2012.
[20] C. Xia, H. Lu, B. C. Ooi, and J. Hu. Gorder: An efficient method for KNN join processing. In *VLDB*, pages 756–767, 2004.
[21] X. Xiong, M. F. Mokbel, and W. G. Aref. SEA-CNN: Scalable processing of continuous k-nearest neighbor queries in spatio-temporal databases. In *ICDE*, pages 643–654, 2005.